# Mechanism of Si doping in Plasma Assisted MBE Growth of $\beta$-Ga$_2$O$_3$

Nidhin Kurian Kalarickal[1], Zhanbo Xia[1], Joe McGlone[1], Sriram Krishnamoorthy[1], Wyatt Moore[1], Mark Brenner[1], Aaron R. Arehart[1], Steven A. Ringel[1,2], Siddharth Rajan[1,2]

[1]Department of Electrical and Computer Engineering, The Ohio State University, Columbus, OH 43210, USA

[2] Department of Materials Science and Engineering, The Ohio State University, Columbus, OH 43210, USA

We report on the origin of high Si flux observed during the use of Si as a doping source in plasma assisted MBE growth of $\beta$-Ga$_2$O$_3$. We show on the basis of secondary ion mass spectroscopy (SIMS) analysis that Si flux is not limited by the vapor pressure of Si but by the formation of volatile SiO. The low sublimation energy of SiO leads to weak dependence of the SiO flux of Si cell temperature and a strong dependence on the background oxygen pressure. Extended exposure to activated oxygen results in reduction of SiO flux due to the formation of SiO$_2$ on the Si surface. The work reported provides key understanding for incorporating Si into future oxide-based semiconductor heterostructure and device MBE growth.

The high breakdown voltage [1] and availability of bulk substrates grown from melt [2-4] makes $\beta$-Ga$_2$O$_3$ promising for various applications, including power switches [5, 6], high frequency amplifiers [7, 8] and high temperature electronics [9, 10]. High quality epitaxial growth with low defect density and a wide range of controllable n-type doping [2-4, 11] are critical enablers for these applications. Variety of growth techniques like molecular beam epitaxy [12, 13], metal organic chemical vapor deposition [14], halide vapor phase epitaxy [15] and low pressure chemical vapor deposition [16] have been utilized to realize high quality epitaxial layers of $\beta$-Ga$_2$O$_3$. Since the conduction band of $\beta$-Ga$_2$O$_3$ is largely made up of Ga s oribitals, group II elements like Si (30 meV), Ge (30 meV) and Sn (60 meV) provide shallow donor levels [9, 17, 18,] and a detailed understanding of the doping process of each one of them is critical in realizing the full potential of $\beta$-Ga$_2$O$_3$ based devices.

Si is the preferred n-type shallow donor in many III-V materials like GaN (20 meV) and GaAs (6 meV) due to its low activation energy and compatibility with epitaxial growth processes. In the case of molecular beam epitaxy, elemental high-purity silicon is typically used, with the effusion cell maintained at high temperatures (typically 1000 °C -1300 °C) required for sublimation of the solid Si. This typically provides excellent control of doping density up to $10^{20}$ cm$^{-3}$ [19, 20]. Plasma and ozone-based molecular beam epitaxy growth have been used to obtain Si-doped n-type $\beta$-Ga$_2$O$_3$ with excellent transport properties [9, 21]. However, the cell temperatures necessary to achieve Si doping have been found to be significantly lower [9, 22, 23] than that used typically for other non-oxide material systems. Since Si has a high sticking coefficient, this suggests that the flux of Si available at a given temperature is significantly higher in the oxide MBE environment than in III-Nitride or III-As growth environments. In this paper we provide an explanation of the mechanism of Si doping in oxygen-based MBE.

The behavior of a conventional effusion cell in an MBE environment is understood through the Clausius-Clapeyron relation [24]

$$P = P_o e^{\left(-\frac{\Delta H}{KT}\right)}, \tag{1}$$

where $\Delta H$ is the latent heat of vaporization, K is the Boltzmann constant, and P is the vapor pressure. This vapor pressure provides a constant atomic flux ($J$) at the substrate growth surface which can be computed at a distance $L$ (cm) away from the cell given by [25]

$$J = \frac{Pr^2}{L^2}\left(\frac{N_A}{2\pi m K_b T}\right)^{\frac{1}{2}} \cos(\phi), \tag{2}$$

where $P$ is the vapor pressure of the source at the given temperature ((1)), $r$ is the aperture of the crucible, $m$ is the atomic mass of source element, $K_b$ is the Boltzmann constant, $N_A$ is the Avogadro's number, $T$ is the cell temperature and $\phi$ (30°) is the angle at which the cell is mounted relative to the substrate. Fig.1 (a) shows the expected molecular beam flux of a standard Si cell (cylindrical crucible with $r \sim 0.7$ cm) as a function of temperature in an MBE chamber ($L \sim 20$ cm). Fig.1 (a) also shows previous reported Si flux values in blue boxes showing clear discrepancy (over four orders of magnitude) with the theoretically expected curves. It is therefore clear that sublimation of silicon, as predicted by (1) cannot explain the flux of Si available during growth.

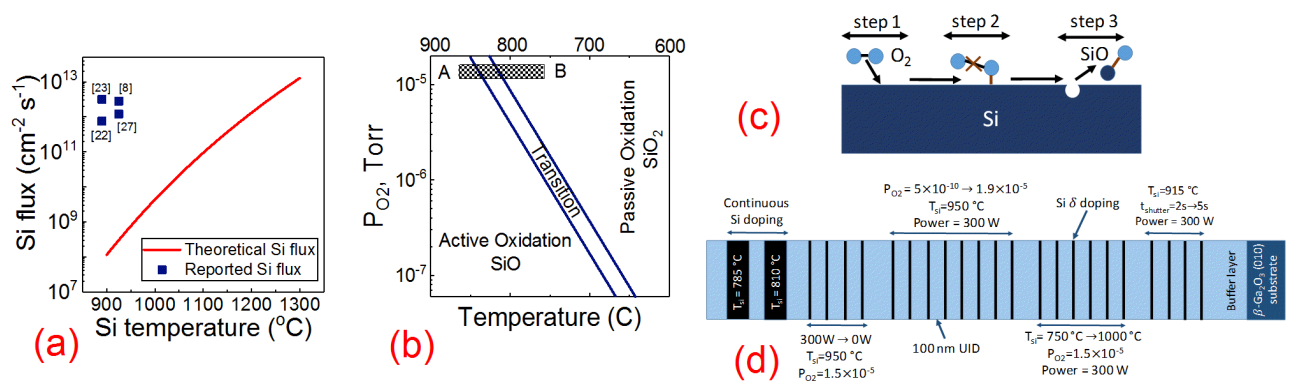

FIG. 1. (a) Theoretically expected Si flux seen at the substrate surface as function of temperature. Vapor pressure data for Si was referred from [26]. (b) Si-O pressure-temperature diagram. (c) Process of active oxidation giving volatile SiO. (d) Epitaxial structure used for SIMS analysis

Since Si is in an oxidizing environment, we consider the Si-O pressure-temperature diagram [28], as shown in Fig.1 (b). In the temperature region of interest there are essentially two oxidation regimes separated by a transition region given by

$$Si\ (s) + \frac{1}{2}O_2(g) \rightarrow SiO\ (g), \tag{3}$$

$$Si\ (s) + O_2(g) \rightarrow SiO_2\ (s), \tag{4}$$

where (3) represents active oxidation and (4) represents passive oxidation of Si. Active oxidation is characterized by the formation of volatile $SiO$ leaving behind a clean Si surface. As shown in Fig.1 (c) this can be understood as a three-step process where the oxygen molecule first gets adsorbed onto the Si surface (step 1) followed by the surface reaction forming an SiO surface molecule (step 2) and finally the desorption of the SiO (step 3). In contrast, passive oxidation proceeds via the formation of a thin film of $SiO_2$ covering the Si surface. The formation of SiO is a unique feature of oxidizing environments, and we show that this is a key step in the anomalously high Si doping flux observed during plasma-assisted MBE growth of $\beta$-$Ga_2O_3$.

To investigate the mechanism of Si doping, experiments were carried out where the cell temperature, background pressure, plasma power, and time were varied during the growth of $\beta$-$Ga_2O_3$. The resulting Si doping density was then measured using secondary ion mass spectrometry (SIMS, Evans Analytical Group [29]). MBE growth of the $Ga_2O_3$ SIMS sample was carried out in a Riber MBE Solutions M7 system equipped with a Veeco oxygen plasma source. Growth was initiated on Tamura (010) Fe doped substrate in slightly Ga rich condition (Ga BEP = $8 \times 10^{-8}$ Torr), oxygen pressure of $1.5 \times 10^{-5}$ Torr, oxygen plasma power of 300 W and a growth temperature of 630 °C (measured using an optical pyrometer calibrated to Si). The epitaxial test structure is shown in Fig.1 (d). A nucleation/buffer layer of 200 nm was used to mitigate the influence of environmental Si contamination from the growth interface.

To study the effect of Si cell temperature on the Si sheet density, the Si cell temperature was varied from 1000 °C to 750 °C (50 °C steps) and at each temperature the Si shutter was opened for 3s. This provides delta doping of a single monolayer of $\beta$-$Ga_2O_3$ since each monolayer takes 3.3 s of growth (2.7 nm/min). Each Si flux point is separated by 100 nm of UID $\beta$-$Ga_2O_3$ layer (Fig.1 (d)). Fig.2 (a) shows the integrated Si concentration for the 3-second deposition cycle (here on referred to as $Si_{3s}$) as a function of temperature. There is a marginal increase in $Si_{3s}$ from $1.12 \times 10^{13}$ cm$^{-2}$ to $1.54 \times 10^{13}$ cm$^{-2}$ as the temperature is increased from 800 °C to 1000 °C corresponding to an increase of 28% in flux over a temperature range of 200 °C, which is significantly lower than the nearly three orders of magnitude change in Si beam flux over the same temperature range (Fig.1 (b)). Additionally, $Si_{3s}$ shows a large reduction as the cell temperature is lowered from 800 °C to 750 °C. This behavior of the Si source can be understood based on the Si-O P-T diagram (Fig.1 (b)). As the Si temperature is lowered from 1000 °C to 750 °C we are essentially traversing the P-T diagram along the line A-B (Fig.1 (b)). This shifts the oxidation characteristics from active to passive, promoting the growth of thermal SiO$_2$ which limits the flux of SiO. D'Evelyn et.al. previously observed that the active oxidation reaction probability increases by 30% between 767 °C and 987 °C [30]. This is very similar to the 28% increase in Si$_{3s}$ observed in this experiment between 800 °C and 1000 °C.

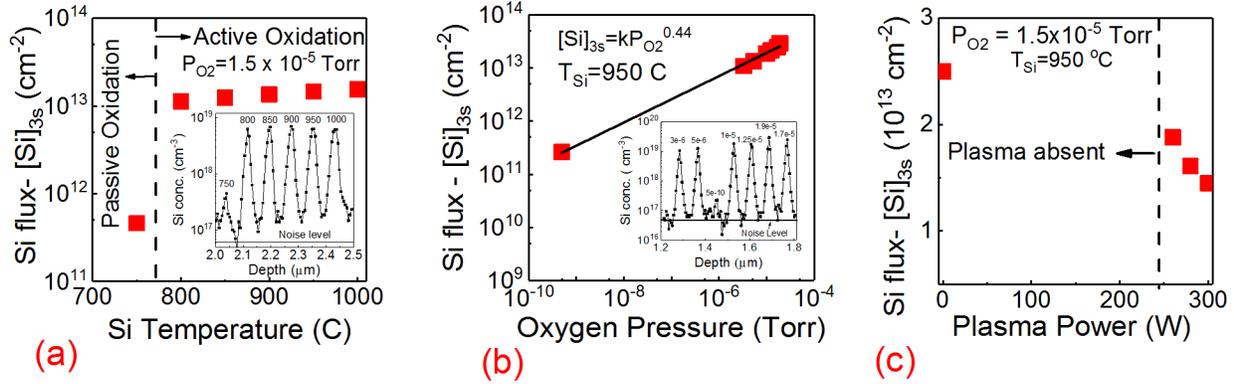

FIG. 2. (a) Si$_{3s}$ vs. oxygen chamber pressure, (b) Si$_{3s}$ vs. Si cell temperature, (c) Si$_{3s}$ vs. oxygen plasma power. Insets show the corresponding SIMS profiles.

To investigate the effect of oxygen chamber pressure on the Si flux seven different oxygen pressures of $3 \times 10^{-6}$, $5 \times 10^{-6}$, $1 \times 10^{-5}$, $1.25 \times 10^{-5}$, $1.5 \times 10^{-5}$, $1.7 \times 10^{-5}$, $1.9 \times 10^{-5}$ and $5 \times 10^{-10}$ Torr were studied with the Si cell at 950 °C. Following every 100nm of UID layer, the growth was interrupted (Ga shutter closed) and the oxygen pressure was altered from the growth pressure of $1.5 \times 10^{-5}$ Torr to a new value. This new pressure was maintained inside the chamber for 10 min followed by which the Si shutter was opened for 3s (Fig.1 (d)). Once the Si shutter closes the oxygen pressure was changed back to the growth pressure, Ga shutter was opened and the growth was resumed. Since the $\beta$-Ga$_2$O$_3$ growth temperature is low (630 °C), all the Si atoms are expected to stick to the epitaxial layer irrespective of whether the growth is on or off. The lowest pressure of $5 \times 10^{-10}$ Torr was obtained by pumping out the chamber for 12 hours till chamber base pressure was reached. Except for the lowest pressure of $5 \times 10^{-10}$ Torr, the oxygen plasma was active at all the other pressure points. For a standard effusion source, the atomic flux density should be independent of the chamber pressure. However, the variation of Si$_{3s}$ as a function of oxygen pressure shown in Fig.2 (b), shows a dependence on the chamber pressure as it is increased from $5 \times 10^{-10}$ torr to $1.9 \times 10^{-5}$ torr. A fit of this data to a power law curve gives a relationship between Si$_{3s}$ and oxygen pressure

$$Si_{3s} = kP_{O_2}^{0.44} \quad , \tag{5}$$

where $P_{O_2}$ refers to the chamber pressure of oxygen in Torr and $k$ ($10^{15.47}$) is a fitting constant. Under equilibrium active oxidation conditions following equation (3), the Si flux and the oxygen partial pressure should follow

$$Si_{3s} = kP_{O_2}^{0.5} , \tag{6}$$

where k corresponds to the rate constant of the overall reaction (step1 + step2 + step3 in Fig.1 (c)) and the exponent of 0.5 is the stoichiometric ratio of the reactants. The fitted value of the exponent (0.44) in this experiment is therefore in close agreement with the theoretically expected rate exponent (0.5) of SiO formation. Therefore considering both the temperature variation and pressure variation of $Si_{3s}$, the anomalous Si flux can be attributed to the formation of SiO.

To understand the effect of activated oxygen on the Si flux, the plasma power was varied from 300 W to 0 W. The source temperature (950 °C) and the oxygen pressure ($1.5 \times 10^{-5}$ Torr) were kept constant during this study. Since the oxygen plasma power determines the partial pressure of activated oxygen species, this series provides insight into the effect of activated oxygen species ($O_2^*$) on the Si flux. The results of this experiment are shown in Fig.2 (c). Reducing the oxygen plasma power results in an increase in $Si_{3s}$ suggesting that activated oxygen suppresses the formation of SiO. Since active oxygen ($O_2^*$) has a much higher oxidation efficiency compared to molecular oxygen, the former can promote the complete oxidation of Si and SiO to $SiO_2$. Therefore an increase in pressure of activated oxygen (increase in plasma power) favors passive oxidation of the Si surface to $SiO_2$ and reduces the flux of SiO.

The effect of shutter open time was characterized using two experiments, one where the shutter time was limited to few seconds and the other where the shutter was kept open for extended time. Fig.3 (a) shows the integrated Si concentration as a function of shutter time (few s). The Si cell temperature and oxygen pressure were maintained at 915 °C and $1.5 \times 10^{-5}$ Torr, respectively. Si sheet density increases linearly with shutter time indicating a constant SiO flux for short shutter pulses.

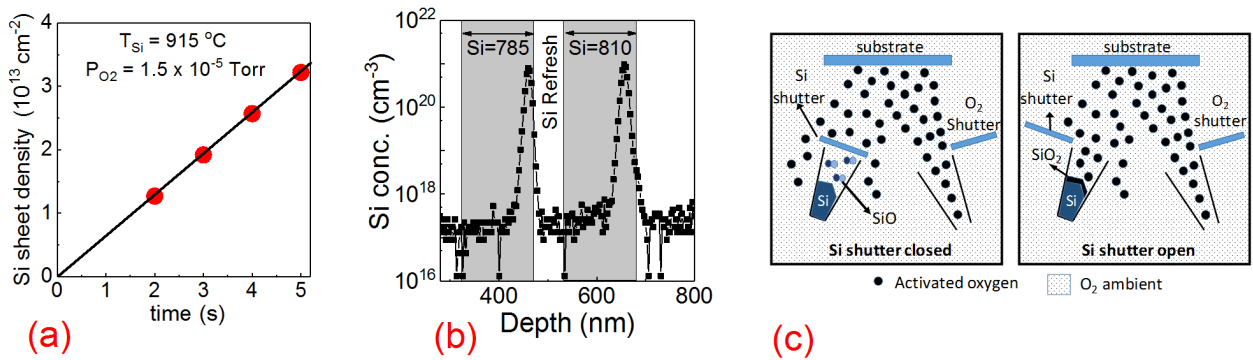

FIG. 3. (a) Si sheet density as a function of shutter open time. (b) SIMS profile showing Si concentration under extended shutter open time. (c) MBE chamber diagram showing the interaction of Si source with molecular and activated oxygen under different shutter conditions.

To study the impact of extended shutter time a 150 nm thick layer of $\beta$-$Ga_2O_3$ was grown with the Si shutter open at two different Si cell temperatures of 785 °C and 815 °C. SIMS profile of this layer (Fig.3 (b)) shows that the Si flux was active only for the first 40 nm after which it diminishes to noise level. This corresponds to a shutter open time of approximately 15 min (2.7 nm/min). This behavior can be better understood by considering the interaction of activated oxygen with the Si cell as shown in Fig.3 (c). Since activated oxygen has a short lifetime, the Si source gets directly exposed to it only when the shutter is kept open. As activated oxygen promotes the complete oxidation of Si to $SiO_2$, extended opening of the shutter facilitates the direct interaction of Si and activated oxygen promoting the growth of $SiO_2$ limiting the flux of SiO.

The oxidized Si surface formed due to extended exposure to oxygen plasma can be cleaned up by a silicon refresh process where the oxidized source is heated to high temperatures (> 1100 °C) causing the disproportionation reaction [28]

$$SiO_2(s) + Si(s) \rightarrow 2SiO(g). \qquad (7)$$

This removes the surface layer of $SiO_2$ leaving behind a clean Si surface.

In conclusion, we have shown through extensive SIMS analysis that the anomalous Si flux observed in the plasma assisted MBE growth of $\beta$-$Ga_2O_3$ arises from the formation of volatile SiO. This was confirmed by studying the variation of Si flux with temperature and oxygen pressure. Reduction in Si cell temperature was found to shift the oxidation condition from active to passive causing large reduction in beam flux. Since an oxidation reaction is at play, the Si flux depends heavily on the oxygen partial pressure and follows a power law relation of order 0.5. Activated oxygen was found to suppress the formation of SiO and promotes the growth of $SiO_2$ on the Si surface. A constant flux vs. time relation was confirmed for short shutter pulses while extended exposures diminish the flux of SiO. A Silicon refresh can be adopted to clean an oxidized Si surface by heating the cell to high temperature (>1100 °C). The work described here can help guide future material and device investigation based on $\beta$-$Ga_2O_3$ and in addition, the mechanism of Si doping is of relevance for any oxygen based vacuum growth environment.

We acknowledge funding from the AFOSR under Grant no. FA9550-18-1-0479 (GAME MURI, program manager Dr. Ali Sayir). We acknowledge funding from The Ohio State University Institute of Materials Research (IMR) Multidisciplinary Team Building Grant.

# REFERENCES

[1] Yan, X., Esqueda, I.S., Ma, J., Tice, J. and Wang, H., *Applied Physics Letters*, *112*(3), p.032101. 2018.
[2] Stepanov, S.I., Nikolaev, V.I., Bougrov, V.E. and Romanov, A.E., *Rev. Adv. Mater. Sci*, *44*, pp.63-86. 2016.
[3] Higashiwaki, M., Sasaki, K., Kuramata, A., Masui, T. and Yamakoshi, S., *physica status solidi (a)*, *211*(1), pp.21-26. 2014.
[4] Tsao, J.Y., Chowdhury, S., Hollis, M.A., Jena, D., Johnson, N.M., Jones, K.A., Kaplar, R.J., Rajan, S., Van de Walle, C.G., Bellotti, E. and Chua, C.L., *Advanced Electronic Materials*, *4*(1), p.1600501. 2018.
[5] Hu, Z., Zhou, H., Feng, Q., Zhang, J., Zhang, C., Dang, K., Cai, Y., Feng, Z., Gao, Y., Kang, X. and Hao, Y., *IEEE Electron Device Letters*, *39*(10), pp.1564-1567. 2018.
[6] Hu, Z., Nomoto, K., Li, W., Zhang, Z., Tanen, N., Thieu, Q.T., Sasaki, K., Kuramata, A., Nakamura, T., Jena, D. and Xing, H.G., 2018. *Applied Physics Letters*, *113*(12), p.122103.
[7] Green, A.J., Chabak, K.D., Baldini, M., Moser, N., Gilbert, R., Fitch, R.C., Wagner, G., Galazka, Z., Mccandless, J., Crespo, A. and Leedy, K., *IEEE Electron Device Letters*, *38*(6), pp.790-793. 2017.
[8] Xia, Z., Joishi, C., Krishnamoorthy, S., Bajaj, S., Zhang, Y., Brenner, M., Lodha, S. and Rajan, S., *IEEE Electron Device Letters*, *39*(4), pp.568-571. 2018.
[9] Higashiwaki, M., Sasaki, K., Kamimura, T., Hoi Wong, M., Krishnamurthy, D., Kuramata, A., Masui, T. and Yamakoshi, S., *Applied Physics Letters*, *103*(12), p.123511. 2013.
[10] Bartic, M., Baban, C.I., Suzuki, H., Ogita, M. and Isai, M., *Journal of the American Ceramic Society*, *90*(9), pp.2879-2884. 2007.
[11] Pearton, S.J., Yang, J., Cary IV, P.H., Ren, F., Kim, J., Tadjer, M.J. and Mastro, M.A., *Applied Physics Reviews*, *5*(1), p.011301. 2018.
[12] Okumura, H., Kita, M., Sasaki, K., Kuramata, A., Higashiwaki, M. and Speck, J.S., *Applied Physics Express*, *7*(9), p.095501. 2014.
[13] Tsai, M.Y., Bierwagen, O., White, M.E. and Speck, J.S., *Journal of Vacuum Science & Technology A: Vacuum, Surfaces, and Films*, *28*(2), pp.354-359. 2010.  [14] Zhang, Y., Alema, F., Mauze, A., Koksaldi, O.S., Miller, R., Osinsky, A. and Speck, J.S., *APL Materials*, *7*(2), p.022506. 2019.
[15] Murakami, H., Nomura, K., Goto, K., Sasaki, K., Kawara, K., Thieu, Q.T., Togashi, R., Kumagai, Y., Higashiwaki, M., Kuramata, A. and Yamakoshi, S., *Applied Physics Express*, *8*(1), p.015503. 2014.
[16] Rafique, S., Karim, M.R., Johnson, J.M., Hwang, J. and Zhao, H., *Applied Physics Letters*, *112*(5), p.052104. 2018.
[17] Han, S.H., Mauze, A., Ahmadi, E., Mates, T., Oshima, Y. and Speck, J.S., *Semiconductor Science and Technology*, *33*(4), p.045001. 2018.
[18] Neal, A.T., Mou, S., Rafique, S., Zhao, H., Ahmadi, E., Speck, J.S., Stevens, K.T., Blevins, J.D., Thomson, D.B., Moser, N. and Chabak, K.D., *Applied Physics Letters*, *113*(6), p.062101. 2018.
[19] Afroz Faria, F., Guo, J., Zhao, P., Li, G., Kumar Kandaswamy, P., Wistey, M., Xing, H. and Jena, D., *Applied Physics Letters*, *101*(3), p.032109. 2012.
[20] Li, W.Q., Bhattacharya, P.K., Kwok, S.H. and Merlin, R., *Journal of applied physics*, *72*(7), pp.3129-3135. 1992.
[21] Chabak, K.D., McCandless, J.P., Moser, N.A., Green, A.J., Mahalingam, K., Crespo, A., Hendricks, N., Howe, B.M., Tetlak, S.E., Leedy, K. and Fitch, R.C., *IEEE Electron Device Letters*, *39*(1), pp.67-70. 2017.
[22] Zhang, Y., Neal, A., Xia, Z., Joishi, C., Johnson, J.M., Zheng, Y., Bajaj, S., Brenner, M., Dorsey, D., Chabak, K. and Jessen, G., *Applied Physics Letters*, *112*(17), p.173502. 2018.
[23] Zhang, Y., Joishi, C., Xia, Z., Brenner, M., Lodha, S. and Rajan, S., *Applied Physics Letters*, *112*(23), p.233503. 2018.
[24] Brown, Oliver LI. *Journal of Chemical Education* 28.8: 428. 1951
[25] Franchi, Secondo. *Molecular Beam Epitaxy*. Elsevier, 2013. p.6.
[26] Desai, Pramond D. *Journal of physical and chemical reference data* 15.3: 967-983. (1986)
[27]Krishnamoorthy, S., Xia, Z., Joishi, C., Zhang, Y., McGlone, J., Johnson, J., Brenner, M., Arehart, A.R., Hwang, J., Lodha, S. and Rajan, S., *Applied Physics Letters*, *111*(2), p.023502. 2017.
[28] Starodub, D., Gusev, E.P., Garfunkel, E. and Gustafsson, T., *Surface Review and Letters*, *6*(01), pp.45-52. 1999.
[29] See https://www.eag.com/techniques/mass-spec/secondary-ion-mass-spectrometry-sims/ for more information on SIMS scan (last accessed July 18, 2019)
[30] D'evelyn, M.P., Nelson, M.M. and Engel, T., 1987. *Surface science*, *186*(1-2), pp.75-114. 1987.